\pdfoutput=1
\documentclass{new_tlp}
\usepackage{mathptmx}
\usepackage{xcolor}
\usepackage{graphics}
\usepackage{graphicx}
\usepackage{hyperref}

\newcommand\bcmdtab{\noindent\bgroup\tabcolsep=0pt%
  \begin{tabular}{@{}p{10pc}@{}p{20pc}@{}}}
\newcommand\ecmdtab{\end{tabular}\egroup}

\newcommand{\hide}[1]{}

  \title[Theory and Practice of Logic Programming]
        {Introducing Computer Science to High School Students through Logic Programming\footnote{To appear in Theory and Practice of Logic Programming}}

  \author[Yuen, Reyes and Zhang]
         {TIMOTHY T. YUEN\\
         University of Texas at San Antonio\\
	\email{timothy.yuen@utsa.edu}
         \and MARITZA REYES \\
         University of Texas at Austin\\
	\email{maritza\_reyes@utexas.edu}
         \and YUANLIN ZHANG \\
        Texas Tech University\\
	\email{y.zhang@ttu.edu}}

\jdate{September 2017}
\pubyear{2018}
\pagerange{\pageref{firstpage}--\pageref{lastpage}}
\doi{S1471068401001193}

\begin{document}

\label{firstpage}

\maketitle

  \begin{abstract}
    This paper investigates how high school students in an introductory computer science course approach computing in the Logic Programming (LP) paradigm. This qualitative study shows how novice students operate within the LP paradigm while engaging in foundational computing concepts and skills:  students are engaged in a cyclical process of abstraction, reasoning, and creating representations of their ideas in code while also being informed by the (procedural) requirements and the revision/debugging process. As these computing concepts and skills are also expected in traditional approaches to introductory K-12 CS courses, this paper asserts that LP is a viable paradigm choice for high school novices. This paper is under consideration in Theory and Practice of Logic Programming (TPLP). 
    
%(?? How about revising the left sentences:  As there are the same computing concepts and skills that are commonly expected in the current standards on CS education for students from elementary schools to high schools. This paper asserts that LP is a viable paradigm choice for novices at high school level. ??)
  \end{abstract}


%%\tableofcontents
\section{Introduction}

%\subsection{What is Computer Science and Why Logic Programming}

%What is computer science, start from computational thinking, then K-12 (progression) 
  
%Computer Science was first introduced to the K-12 (??) education as early as in 1960s (e.g., Logo programming for children by  Feurzeig, Papert and Solomon \cite{abelson1974logo}).   

%There is recent consensus that Computer Science should be taught to all students in K-12 \cite{smith16computer}. 

%A central question to Computer Science education in K-12 is what to teach and why. Recently, K-12 CS education has been studied under the notion of computational thinking \cite{allan2011compytational,barr2011bring,brennan2012new,wing2011research}. Though there have been many efforts in defining  computational thinking is \cite{grover2013computational}, more effort is needed to have a full understanding of computational thinking\cite{hemmendinger2010a,csta2012,voogt2015computational}. As shown in the recent curriculum standards and framework (e.g., Computer Science Teachers Association Standards \cite{csta2012} and K-12 Computer Science Framework \cite{csFramework16}) and the AP course (a high school course in USA) ``Computer Science Principles" \cite{apcs}, CS education must address the important CS topics of  {\em abstracting}, {\em algorithms}, {\em programming} and {\em communicating}. Both object-oriented (OO) and procedural paradigms have been used in teaching introductory computer science courses \cite{force2013computer}.

The debate in paradigm selection in introductory computer science (CS) courses is often split between object-oriented (OO) and procedural paradigms, which also leads to discussions on programming language choice. This choice is difficult as there are no standard languages or paradigms to use in the field. Though both these approaches have been successful in introductory CS courses, some research has also shown only minimal differences when comparing the outcomes between paradigms \cite{chen2006relationship,vilner2007fundamental}. However, research on the teaching of introductory CS courses also reveals the limitations of those approaches \cite{pears2007survey,chakravarty2004risks} and that the 
suitability of Java, C/C++ for education is debatable.
% and as objects-first. 
Often, these discussions of introductory courses are aimed at the university level. However, high school is a critical juncture when CS educators have one last opportunity to engage students in CS and provide adequate preparation before they select their majors and begin university. Thus, this paper emphasizes the importance of introductory CS courses, as well as paradigm/language selection, at the high school level. At this time, more research is needed on how to best teach CS in high schools.

Some research shows that teaching programming to middle and high school students is challenging \cite{sherin1993dynaturtle}.  
By a recent survey on teaching CS in K-12 using programming 
(K-12 is a USA school system including Kindergarten, first grade (usually 6–7 years old) until 12th grade. Elementary schools usually include 1st to 5th grade, middle schools 6th to 8th grade, and high schools 9th to 12th grade),   
the majority of education research uses visual programming languages \cite{lye2014review}. Examples of these languages include Scratch \cite{resnick2009scratch}, Toontalk \cite{kahn2004toontalk}, and Alice \cite{cooper2000alice}.
% \cite{resnick2009scratch} 
With these visual languages, students build programs using graphical objects and drag-and-drop interfaces, which significantly reduces the challenges for students to learn the syntax (compared to text-based programming). Another prominent feature of these environments is to encourage and  facilitate  tinkering and fit the needs of students who prefer {\em tinkering} to {\em logic and planning} \cite{papert1980mindstorms,resnick2013designing}. On the one hand, these languages and environments have been very successful in reaching a large K-12 population. On the other hand, more research is needed to understand how computation thinking occurs as students are tinkering while using visual programming languages \cite{guzdial2004programming,lye2014review}. While substantial efforts have been made in introducing CS to younger audiences through visual programming languages which are largely procedural or OO based, logic programming (LP) based approaches have been largely ignored by the K-12 Computer Science education 
research community \cite{pears2007survey,lye2014review}, despite continuing efforts in both teaching and research on 
LP.  

\subsection{Logic Programming and CS Education}
Prolog may be the most well-known instance in which LP was used to teach CS. In particular, Prolog was used to teach children in the 1980s \cite{kowalski1982logic,kowalski1987logic,nichol1988prolog,guzdial2004programming}.
% \cite{kowalski1987logic,nichol1988prolog}
As a pioneer, Kowalski \citeyear{kowalski1987logic} focused on the declarative aspects of Prolog and restricted the use of procedural aspects of Prolog to a minimum. Later, researchers and practitioners found that the procedural aspects of Prolog have been the main source of misconceptions and difficulties \cite{mendelsohn1990programming}, while the benefits of its declarative aspects were acknowledged.
In the last two decades, Prolog has made some appearances: listed in high school curriculum \cite{scherz1995logic}, and taught to gifted and talented high school students \cite{stutterheim2013forty} as well as to general high school and undergraduate students \cite{levesque2012,beux2015computational} though it does not enjoy the same attention as procedural and OO paradigms.
In the last two decades, a breakthrough in LP research was the establishment 
of the Answer Set Programming (ASP) 
\cite{brewka2011answer,gelfond2014knowledge,kowalski2014logic}, a paradigm which has 
inherited the declarative nature of Prolog while fully removing its 
procedural features. 

%We will argue below that ASP provides an environment suitable for teaching Computer Science in middle and high schools. 
ASP meets the expectations of
a programming environment for young children
\cite{papert1980mindstorms,guzdial2004programming,resnick2009scratch,grover2013computational}:
``low floor" (easy to get started) and  
a ``high ceiling" (opportunities to create increasingly
complex projects over time).
ASP has a simple syntax. LP, in general, and ASP, in particular, are closer to natural language compared with other programming paradigms, allow natural modeling of problems and thus allows the focus on the essence of the problems. Also, ASP is a full-fledged  programming language that can be used to solve NP complete problems or problems in a higher level of the complexity hierarchy. 

As for CS education at the level of both K-12 and colleges, significant efforts in the last two decades 
were made to understand what CS is and what to teach about CS.
One line of work is carried out under the notion of Computational Thinking
\cite{allan2011compytational,barr2011bring,brennan2012new,wing2011research}. Progress is being made on what Computational Thinking is \cite{grover2013computational}, 
but significant effort is still needed to have a full understanding of 
Computational Thinking \cite{hemmendinger2010a,csta2012,voogt2015computational}.
The study of computational thinking has led to the development of several national and international standards for teaching computational thinking at the secondary school level.

As shown in the recent curriculum standards and framework (e.g., Computer Science Teachers Association Standards \cite{csta2012} and K-12 Computer Science Framework \cite{csFramework16}) and the AP course (an Advanced Placement high school course in USA) ``Computer Science Principles" \cite{apcs}, CS education should cover the following components:  {\em abstracting}, {\em algorithms}, {\em programming} and {\em communicating}. 

ASP provides an environment for teaching these components. The objects, variables and relations underlying ASP provide a rich structure for students to practice {\em abstraction} when modeling a problem. For example, students can practice abstraction from the very basic level involving a relation among concrete (constant) objects, to a higher level involving queries containing variables,  and to the fullest level involving rules. There are also many finer abstraction levels in between those three levels. 

As a well-established programming paradigm, ASP 
offers a setting for students to learn and practice all aspects of {\em programming}: design the model (program), edit the program,  and learn the (informal yet rigorous to a great extent) syntax and semantics, recursion, coding and debugging. ASP allows a modeling methodology which starts 
from the description of knowledge (in the problem of concern) in natural language  (e.g., English)
and may need continuous refinement of the English description. 
The transition from English description to ASP rules
provides scaffolds for students to understand both rules and English description in a more precise way. 
Therefore, ASP based modeling 
trains students to practice {\em communicating information}.

As for {\em algorithms}, ASP can cover recursion. The training of students in abstraction and rigorous representation through LP 
is expected to increase students' capacity for 
specifying problems accurately, which is taken as desirable before studying algorithms \cite{kleinberg2006algorithm}.  However, as a pure declarative programming paradigm, ASP is hardly the best option for teaching algorithms. On the other hand, by skipping the need to know algorithms, ASP significantly reduces the gap between (interesting) problems and their computer modeling, which seems to serve better the initiative \cite{smith16computer} in the United States to teach computer science to 
{\em all} students in K-12. 
% the distance from the detailed treatment of algorithms, 
See Section~\ref{sec:limitations} for a 
more detailed discussion.

\subsection{Advantages of Logic Programming}
From the existing work and practice in LP and general programming \cite{pears2007survey,chakravarty2004risks,scherz1995logic,ball2015teach}, the main advantages of LP as related to computing education are: 1) simple syntax and intuitive/declarative semantics; 2) natural connection to abstraction, logical reasoning and knowledge representation which form the foundation of computing and other disciplines; 3) early involvement of non-trivial, interesting and challenging problems (e.g., Sudoku problem); and 4) the mathematical flavor of LP. LP also provides a rich context for embedding essential computing concepts and skills (as argued in the previous subsection). As for industrial relevance, LP and Declarative Programming (DP), in general, have seen their profound application and impact in database query languages, problem (formal) specification languages, and domain specification languages including popular web application development languages such as HTML, CSS and XSLT etc. Currently, ASP is a dominating LP formalism in knowledge representation. 

%A breakthrough in LP research in the last two decades is the establishment of the Answer Set Programming (ASP) \cite{kowalski2014logic,gelfond2014knowledge}, a paradigm which has inherited the declarative nature of Prolog while fully removing its procedural features. 

\subsection{Research Question}
Given the advantages of LP in CS education, 
% and its need in some computing fields, 
more investigation is needed in how LP could be introduced to 
K-12 students. 
%novices. 
Although there is progress being made to expand LP to secondary schools[e.g., \citeN{dovier2016reasoning}], research on how ASP facilitates student learning in introductory computer science is still absent. Therefore, the research question for this study is: \textit{How does adopting ASP to teach an introductory CS course for high school students impact their understanding of computer science and computing?
}

\section{Methods}

This study adopted a qualitative approach to investigate, in-depth, how high school students obtain understanding of CS and computing through LP, specifically ASP. 
\subsection{Setting and Participants}
Participants were recruited from the TexPREP program at Texas Tech University during Summer 2015. TexPREP is a summer enrichment program for 6th-12th graders and offers full day courses in science, engineering, math, and computer science over several weeks. Students in these grades typically range from 11 to 18 years old. Programs such as TexPREP are often used to motivate students to attend university after the 12th grade. 

TexPREP courses are typically taught by local school teachers as well as university faculty members and students. In the case of this study, the class was taught by the third author with some TAs (the second author was also a TA). TexPREP is a selective program: students had to have an A/B+ average, letters of recommendations, a written essay, and a personal interview for consideration. Prospective participants were in their fourth year of TexPREP taking Computer Science 4 (TexPREP-CS4). They had already taken TexPREP-CS2 with Scratch \cite{resnick2009scratch} and TexPREP-CS3 with Alice \cite{kelleher2007using}. The TexPREP-CS4 course teaches the foundations of computing using ASP \cite{gelfond2014knowledge}. 
This course emphasized problem solving and knowledge representation through the ASP paradigm. The course met Monday through Friday from 1:00pm to 1:50pm  from June 17  to July 9 of 2015. 

Informed consent was obtained from both the participants and their parents prior to the start of class. There were sixteen (N=16) participants for the study. There were seven females and nine males. Participants were students at high schools in the Texas Tech area in Lubbock, TX. In terms of rising grades, there were eleven 10th graders, three 11th graders, and two 12th graders. 

\subsection{Course Organization}
The teaching methodology employed is influenced by  the existing work on teaching Logic Programming, e.g., the work by Kowalski \citeyear{kowalski1987logic},
Sterling and Shapiro \citeyear{sterling1994art},
Clocksin and Mellish \citeyear{clocksin2003programming}, and
Gelfond and Kahl \citeyear{gelfond2014knowledge}.
Course topics were centered around building computer models, using objects and relations, for solving problems. This section describes how course activities are structured with respect to problem-solving methodologies within the LP paradigm.

\subsubsection{SPARC}

This course approached CS through SPARC\footnote{The manual for SPARC 
is available at \url{https://github.com/iensen/sparc/blob/master/User_Manual/Sparc_Manual.pdf}. It contains detailed information on the 
installation and use of a SPARC solver.} \cite{balai2013towards}
which is the result of introducing sorts into ASP.
A SPARC program consists of three sections: {\em sorts section} where 
sorts are defined, {\em predicates section} where the sorts of the 
parameters of each predicate is declared and {\em rules section} 
which consists of ASP rules. A full example is given in Section~\ref{sec:modelingMethodology}. 

We use ASPIDE \cite{FebbraroRR11} as the 
programming environment for SPARC. ASPIDE \footnote{
ASPIDE is designed with advanced projects and users in mind. 
Its user interface is quite involved. 
Also the installation and maintenance of 
ASPIDE, a stand alone
software, are a challenge in our teaching \cite{MarcopoulosRZ17}. 
In our more recent outreach to middle and high schools,
we use onlineSPARC -- an online programming environment
for SPARC with simple and only necessary functionalities including an editor, query asking, 
obtaining answer sets and rendering literals in answer
sets that represent a drawing or animation \cite{MarcopoulosRZ17}. 
}  provides a
graphical user interface for students to edit a program,
ask queries and get answer sets of the program.

We choose SPARC,  
instead of popular ASP languages such as DLV
\cite{alviano2011disjunctive} or Clingo \cite{gebser2011potassco}, as our
teaching language for the following reasons.
First, the three-section structure provides a 
direct support of the modeling methodology (see next subsection).
Second, the introduction of sorts helps
avoid thinking about/teaching safety conditions 
on rules. Finally, sorts also help provide 
warning on type errors so that programmers can 
identify bugs early. 

\subsubsection{Explicit Modeling Methodology}
\label{sec:modelingMethodology}
Before the discussion of any modeling, we always first give the \textbf{problem description}, which consists of two components: the \textbf{knowledge in the problem} and the \textbf{questions and intended answers}.  

Here is an example of a problem description:
% through the Family Problem:
\begin{itemize}
\item \textbf{Knowledge in the problem}: There is a family. John is the father and Joan is the mother. The children are Jim, Bill, and Sam.
\item \textbf{Questions and answers}: Is John the father of Bill? Yes. Is John the dad of Bill? Yes. Is John a parent of Bill? Yes. 
\end{itemize}

Students can be easily involved in understanding the problem, giving answers to the questions,  as well as  extending the knowledge and questions. 

Note that we emphasize both questions and intended 
answers in problem description. Usually the knowledge 
in a problem description (particularly in a 
real life setting) is not complete or contains 
extra information. The questions and intended 
answers help the students to understand/decide 
the relevant knowledge to represent. 
The students are expected to understand what 
knowledge (and the reasoning process) is needed to
obtain the intended answers. However, 
students are not encouraged to focus on ``algorithmic'' ideas to ``find'' a solution to a 
question. For example, to model the Sudoku problem (or word problems in  algebra or physics), students are expected to use intended answer 
to help them to clarify knowledge needed
and to ``debug" if they miss any knowledge
or use wrong knowledge. But students are not 
encouraged to figure out how to construct a final concrete answer 
manually or systematically. (In fact, a more specific modeling 
methodology for constraint satisfaction problems 
is needed for students to model the Sudoku problem.)

Students are expected to build computer models for the problems so that the questions in the problems can be answered by the computers correctly.
The next part is a two-step explicit  methodology to model a problem:

\begin{enumerate}
\item Identification.
	\begin{enumerate}
		\item \textbf{Objects in the problem}. Examples for the problem above include John. 
		\item \textbf{Relations in the problem}. Examples for the problem above include the dad and father relation between two persons. Students are expected to write the relation in the atom form with compulsory comments on its meaning. E.g., \texttt{dad(X, Y)} denotes that person \texttt{X} is the dad of person \texttt{Y}, and
\texttt{father(X, Y)} denotes that person \texttt{X} is the father of person \texttt{Y}.  
		\item \textbf{Knowledge in the problem}. E.g., John is the father of Bill. Students are expected to list all knowledge (explicit in the problem description or implicit in our common sense) needed to answer the intended questions. 
	\end{enumerate}
\item Translation/coding.
	\begin{enumerate}
		\item \textbf{Objects} are defined and grouped into sorts. For example, 
		
		\texttt{\#people = \{john, bill, joan, sam\}.} 
		
		where \texttt{\#people} is the sort name while the right side of = employs a standard set notation. Intuitively, the sort\texttt{ \#people} means the set of the four persons. 
	\item \textbf{Relations} are declared. For example, 
	
	\texttt{dad(\#people, \#people)}.
	
	 declares that \texttt{dad} is a relation over \texttt{\#people}.
	\item \textbf{Knowledge} will be translated into rules. For example, the knowledge 
	
	``for any person X and Y, X is the dad of Y if X is the father of Y" 
	
	can be translated into a rule: 
	
	\texttt{dad(X, Y) :- father(X, Y)}. 
	\end{enumerate}
\end{enumerate}

\noindent Here is an example of a full SPARC program: 
\begin{verbatim}
sorts
    #people = {john, bill, joan, sam}.
predicates
    % father(X, Y) denotes that X is the father of Y
    father(#people, #people).
    % dad(X, Y) denotes that X is the dad of Y
    dad(#people, #people).
rules
    father(john, bill).
    dad(X, Y) :- father(X, Y). 
\end{verbatim}
% \todo{I made some tweaks here and there. Please verify that everything is correct.}
In this course, students have a lot of opportunities to identify relations and their meaning (e.g., \texttt{dad(X, Y)} and its meaning) and to figure out what relation is used to represent a piece of knowledge (close to the declaration/specification of a function in an imperative language and the use of a function to solve a problem component). Once they are able to switch freely between the atom form of a relation and the English description of the relation, they are ready to write rules. Students are encouraged to always start from English description of the knowledge before writing any rules. 

\subsubsection{Course Content}
The sequence of major topics covered by this course is: an introduction to Computer Science, modeling methodology, 
use of ASPIDE, applying methodology to model the family problem 
(facts on relations {\tt father} and {\tt mother}, rules on relations of {\tt dad} and {\tt parent}), queries to SPARC programs, reasoning (informal but in a manner as rigorous as possible) with facts and rules in the context of the family problem,
writing rules through iterative refinement of English description,
default negation (only in representing the closed world assumption), modeling the family problem with more relations such
as {\tt sister} etc., modeling the simplified battle field problem 
(a game used in middle school math teaching) and map coloring
problem, recursion (defining one person is 
an {\tt ancestor} of another using 
the {\tt parent} relation), modeling a map problem
(about the neighbor of a state and the color of a state), 
and modeling the map coloring problem and Sudoku problem. Details of the course content, except for the Sudoku problem, can be found in the lecture notes available online \cite{zhang2015}. %There are several comments about our course design. 

\subsubsection{Course Design}
This section now discusses the course design. First, we explain the iterative refinement based methodology used to help students to develop rules from their English description of knowledge they generated for a given problem. The class starts from facts. Students will practice writing 
English sentence(s) about simple facts. These sentences will 
then be translated into facts in SPARC. 
Depending on the problem description, this task might not be trivial.
Guidance and examples are given to help students to get used to 
the methodology for writing facts. 
Next, we introduce rules with body using {\tt dad}. We started from 
students' English description of the relationship between {\tt dad} 
and {\tt father}. The first version of the English description is 
usually hard to translate into a rule using the relation {\tt dad(X, Y)} and {\tt father(X, Y)}. The difficulty arises 
from the flexibility of natural language. Here, English allows us to employ an object and a relation related to the object. An example description of {\tt dad} could 
be ``Dad is the same as father.'' Instead of giving an idiomatic 
representation of such knowledge in ASP, we encourage students
to apply the {\em 
iterative refinement methodology}, a fundamental method in computer science, to refine their description until a rule can be written to 
their satisfaction. In the iterative refinement, students will 
recursively use our problem modeling methodology to identify the 
objects and relations in their description. Students will be asked
what objects are involved in the description above and how they are 
related. With some hints, they can figure out the objects (implicit) 
in the description and relate them using the intended relation 
{\tt dad} and {\tt father}.  This practice is expected to enhance students' reasoning capacity, clarity and preciseness in communication. Consider a refined description of grandparents:
$X$ is a grandparent of $Y$ if $X$ is a parent of a parent of $Y$. This English description involves the similar phenomenon in the example for dad. 
The second occurrence of parent is used to denote both an object and 
parent relation between this object and $Y$. By making the object explicit using a variable $Z$, the description can be further
refined as: $X$ is a grandparent of $Y$ if (there is a person $Z$ such that) $X$ is a parent of $Z$ and $Z$ is a parent of $Y$. 
The translation of the refined description to a rule is now straightforward.
We would also emphasize that our teaching materials are
organized surrounding the modeling methodology. 
The methodology is practiced repeatedly in the family problem
and its extension, the map problem, 
the simplified battle ship problem, 
the map coloring problem and Sudoku problem. 

In the Sudoku problem, we were able to demonstrate the {\em top down designing approach}, another important problem solving skill in
CS, when defining two squares in the same 3 by 3 block. Belt concept is introduced to define the same 3 by 3 block relation: two squares are in the same 3 by 3 blocks if the rows of the squares are in the same {\em belt} and the columns of the squares are in the same {\em belt}.  We next define what is a belt. Two numbers are in the same belt if they are in first 3 numbers (i.e., 1 to 3), in the second 3 numbers (i.e., 4 to 6), or in the third 3 numbers (i.e., from 7 to 9). The key in the top down approach is 
to introduce and use a relation(s) as needed and define this relation(s) later. 

Finally, as an introductory course to CS for high school students, the formal definitions of terms, rules or 
semantics (usually covered in most textbooks for college or graduate students) 
are not covered. However, by formulating definitions (e.g., for parent) in English and translating them into rules, the students are expected to experience and appreciate the rigor in representing the knowledge needed to solve a problem. Given the precise form of the rules and intuitive logical reasoning, the students are also expected to exercise their reasoning capacity to a great extent. 

\subsubsection{Lab Assignments}
Four labs (closely related to topics discussed during class) were given, usually along with a program stub. However, students had the opportunities to modify and/or add: sort definitions, predicate declarations and rules. The first lab was based on the family problem and focusing on the basics of the problem description, modeling methodology and the programming in ASPIDE (for SPARC). The students were asked to define the relations of {\tt father} and {\tt mother} that were discussed during class and any new family relations they were interested in and model them in SPARC. They were also expected to submit a report containing the problem
description and the result from the identification step of the problem
modeling methodology.
In the second lab, students were asked to
define three specific family relations: {\tt mom(X, Y)}, {\tt uncle(X, Y)} 
and {\tt sister(X, Y)}. The third lab is the standard map coloring problem, and the fourth lab is the standard Sudoku problem. 

\subsection{Data Collection}
Three main types of data were collected. Surveys were used to gather background information on what students learned before and after the class. The main data source came from clinical interviews which were used to examine students' conceptual understanding. Scores from lab assignments and the task in the clinical interview were then used to support the qualitative findings. 

\subsubsection{Surveys}

Participants responded to surveys with open-ended questions at the beginning (pre) and end (post) of the course regarding their experiences with computing and declarative programming. The purpose of the surveys was to provide some background information on participants' CS knowledge and to triangulate the findings from the clinical interviews. The survey refers to declarative programming (DP), rather than LP or ASP even though this study focuses on the latter two paradigms. The rationale for referring to the more general DP is because: ASP belongs to the LP paradigm and LP is one paradigm of DP. So, knowing other DP paradigms (such as functional programming) may give an edge to the students in learning ASP. 
The pre-survey asked about participants' general knowledge of computer science and declarative programming. The pre-survey questions were:
\begin{itemize}
\item What is computer science?
\item Describe what you know about declarative programming.
\end{itemize}
The post-survey asked about participants' knowledge of computer science and the topics of this course. The post-survey questions specifically asked what participants learned in this course and how they understood the concepts; that is, rather than asking participants directly to explain declarative programming, these questions asked how they perceived declarative programming:
\begin{itemize}
\item What is computer science?
\item What have you learned, if anything?
\item After having taken this course do you feel that you understood the subject of this course and the tasks that were asked of you? Why or why not?
\end{itemize}

\subsubsection{Clinical Interviews}
Clinical interviews are task-based activities in which the researcher attempts to explore participants' understanding and cognitive process through active questioning and probing \cite{ginsburg1997entering}. Clinical interviews were conducted one-on-one by the research team and the participant. During the 
interview in the last week of class, participants were asked to work out one class-based problem while thinking aloud. 
% Since this was a class assignment during a limited time with few researchers, participants only worked on part of the assignment. 
Researchers asked probing questions in order to prompt participants to explain and clarify their thought processes. All researchers received the same training from the first and third author on conducting the clinical interviews. Clinical interviews lasted 30 minutes or until the participant completed their task -- whichever came first. Clinical interviews were video recorded and transcribed.

We have collected interviews for 16 participants. 
All the other participants were out of town in the last week 
and thus no interviews can be done with them. 

In the interview task, each student was asked to write a SPARC program to represent the knowledge about family relationships. This problem was an extension of the second lab and not a formal assignment in the course.  Some relations, such as \texttt{father(X, Y)}, denoting \texttt{X} is the father of \texttt{Y}, and \texttt{mother(X, Y)} are given. A program stub including the declarations of these relations were also given. The students were asked to write rules to represent the knowledge: 1) Jon is the father of Matthew, and 2) to define the following relations:\texttt{ grandparent(X, Y)}, \texttt{son(X, Y)}, \texttt{aunt(X, Y)}, and \texttt{descendant(X, Y)}. 
%SPARC is a programming language and system instance of ASP paradigm \cite{balai2013towards}. It offers a type system to overcome some challenging syntax restrictions in existing ASP systems, such as DLV \cite{alviano2011disjunctive} or Clingo \cite{gebser2011potassco} and help discover some programming errors early.
The results of the surveys and interviews are presented in the following sections.

\subsubsection{Programming Assignment Grades}

Grades for programming assignments were collected by the instructor of this course and used to support the qualitative findings of this course. Lab assignments were graded by a PhD student (called {\em the grader}) who was not a part of this study. The rubric graded projects for: completion of requirements, abstraction, correctness, coding style/readability, and documentation (commenting) on a 4-point scale.  Completion refers to the extent in which a participant implemented all of the assignment's requirements. Correctness refers to the program's ability to produce the correct results for all test cases created by the teaching staff. Abstraction refers to the extent that participants were able to write generalized code that is reusable and adaptable. Participants' source code was graded on coding style in terms of organization, readability, and adherence to class coding style conventions. Documentation refers to the amount and quality of documentation (comments) that describes the source code. 
% \todo{verify that these are the correct definitions for each rubric category}

\section{Survey Results}
Survey responses were transcribed and analyzed by the researchers using an inductive coding approach in which responses were coded with descriptive labels \cite{corbin2008basics}. That is, the researchers attempted to understand the meaning of each fragment and assigned a descriptive label. Similar labels were merged into larger categories. The main unit of analysis was a sentence. Longer sentences that included multiple ideas were sometimes separated for further coding. Similarly, some consecutive sentences were grouped together for analysis when they continued the same line of thought. 

Through this analysis, researchers found common themes across participants. Responses were grouped by common themes (or labels). The most frequent themes are presented below. Participants were given pseudonyms in the presentation of the results in the form of P\#.

\subsection{What Students Know about Computer Science}
Since this course was the third in the pre-engineering program sequence, participants already had two courses in computing and other courses in science and engineering. Most participants' definition of computer science in the pre-survey centered on the creation of a program that does some task with the programming language being the means through which those tasks are completed [e.g., \textit{The science behind how a computer functions and performs tasks} (P1); \textit{Computer science is using different computer “languages” to get computers to perform tasks} (P13); \textit{Understanding and being able to reproduce the code and programs that make a computer do its task} (P4)]. There was a strong connection between computer science and programming with ten participants mentioning programming or the creation of software in their responses. Some participants acknowledged the importance of computers and technologies in the advancement of society [e.g., \textit{The study [of] how to develop computers to help the world be an easier and more efficient place} (P10); \textit{It has become extremely prominent in our daily lives and must always be improved for technology to come} (P11)].

At the end of the course, participants were asked the same question. In the post-survey, only three participants mentioned programming in their definition of computer science. The participants' responses could be categorized as defining computer science as problem solving [e.g., \textit{Computer science is a study that involves mathematics and helps us understand the way a computer thinks} (P3); \textit{Computer science is the study on how to make a computer work/think like a “human”} (P10); \textit{It is how we get computers to solve problems and how we can use them for simplifying, numbers, solutions, or problems themselves} (P11)] and the study of technologies [e.g., \textit{The science behind a computer and how it functions} (P1); \textit{Computer science is the study of how and why computers work} (P16)]. Thus, by the end of the course, the participants' definition of CS had broadened a bit away from just programming; rather, it become more holistic and more on emphasizing problem solving in order to complete tasks and how technologies can support those processes. 
%(??note this students' view don't change much from pre-survey , see above on P1--this should be okay, I updated this sentence??)
\subsection{What Students Know about Declarative Programming}
In the pre-survey, most participants did not know what DP was; however, many were familiar with the more “traditional” procedural programming and/or had prior programming experience. Eight participants said that they did not know what declarative programming was. Some participants offered an explanation of what declarative programming could be [e.g.,\textit{ I would guess that it is a branch of computer science that has to do with making statements and receiving some form of output from the statements} (P12); \textit{I believe that you give a description to of an object then the program runs it to find the item similar to what was described} (P8)] or explanation that they have done programming before [e.g., \textit{In school, I have learned a little about Java and how Java is a declarative language, Other than that, I don't know much else about declarative programming} (P3)]. In such explanations, participants were attempting to make connections to previous programming experiences, even though those were incorrect.

The remaining five participants have some  understanding of what DP was:
% Rather, they offered some incorrect or incomplete definitions: 
\textit{Declarative programming is stating a statement and being able to program that statement. Giving the computer a “command” and programming it to accomplish that particular command (P5); It is a program where it does the things you type out} (P10); \textit{Declarative programming is more of an elaboration off of algorithmic programming as it is developed for a broader use compared to Algorithmic programming. It is much more elaborates and is well suited for situations with similarities} (P12); and \textit{It is the type of programming needs to be defined before any output action can be achieved} (P15). Participant 16 was the closest to having a correct answer: \textit{I know that declarative programming works like telling a computer rules for what something is then having the computer rules for what something is then having the computer go find all the somethings [sic] in a set.} 

From the participants' responses,  they did not have experience with DP although have had some intuitive understanding of what DP may be. 
% in other paradigms, which included object-oriented, procedural, and imperative.

\subsection{What Students Have Learned in This Course}
\label{sec:whatStudentsLearned}
The survey data showed that most students reported learning how to program with the SPARC language within the context of their lab assignments. Though most did not explicitly articulate their understanding of LP, as they did with procedural-type languages from prior CS experiences, they were able to report how they solved problems using ASP methodologies and tools. Although the participants emphasized learning how to program with SPARC, eleven responses implied that the programming language provided the structure that required students to think about their problems through a deeper, more thorough perspective. P12's response was representative of how learning the programming language helped guide declarative understanding: \textit{I've learned how to reevaluate my thinking process and consider the basis of my thought (so I can program). I've learned how to ‘translate' – by writing my goals/rules/knowledge in English, then transferring and modifying it to fit SPARC code.}

Similar responses include: \textit{I learned how to look at a problem in a different way} (P2); \textit{I have learned how to sort objects, define relationships with predicates, use multiple variables, use multiple predicates/ rules, use tex***\footnote{The words in the original writing were not readable.} rules with variables, and use if when for rules to help simplify solutions} (P11); \textit{I have learned a new coding language to go along with the others I know, which is very valuable to me. I have also learned a new way to look [at] relations between objects} (P15); \textit{New syntax - A new way to specify relations} (P9). These examples show that the rules and structure (i.e., syntax and semantics) that SPARC required and the associated methodologies helped guide participants in their problem solving. Four participants emphasized learning about the syntax and coding style. P8 also mentioned that this was the first course in which she had to type out her programming rather than through a drag-and-drop interface. 

\subsection{How Students Understood the Course Topics and Tasks}
Eight participants stated that they understood the subject of this course within the confines of the course activities and three were indifferent. They understood that there was more to learn with respect to the programming language [e.g., \textit{I think if given more time and work on the computer, I could use this language in the future} (P3); \textit{The prompt and things that were asked were clear and easy to understand. They were easy to pick up on and gave a general understanding – making it relatively easy to complete a given task} (P5); {\em I understand the subject because it was easy. I'm pretty sure that if you wanted to program more complex things, it would be harder} (P13). 

Two participants stated they were able to understand the subject of the course, but were unsure how it was applicable in a real world or everyday setting: \textit{I understand how, but not why. So, I don't understand the purpose of this type of program and how it will be beneficial to the world, other than artificial intelligence, but I feel as though this program requires more to get to that point} (P9) and \textit{I do feel I understood the subject of the course as I did actually learn something. However, I did not always understand exactly what I was supposed to do and I was occasionally completely lost} (P4). Therefore, most participants knew that there was more to learn about LP beyond the contexts with which they were presented. Both P10 and P14 felt that the course was too repetitive with the simple topics and was not challenging enough. 

\subsection{Student Feedback}

Participants were asked to provide feedback on what they thought of the course and its LP activities, and 13 of them responded. Participants seemed to appreciate\footnote{Some feedback overlapping with the positive comments reported in Section~\ref{sec:whatStudentsLearned} is not repeated here.} the different approach that LP/ASP takes to programming as most of them have worked with either Java and/or a visual programming language like Scratch:
\begin{itemize}
\item {\em I like this program because it makes you look at code in a different way. ASPIDE is such a different set of code then the standard JAVA. I like the format of this code and the different symbols.} 
\item {\em I learned a lot about declarative programming and how to use it. The course was actually writing code and not drag and drop.}
\item {\em ...this course has a lot of potential and if executed right it can be very beneficial because the concept is cool and it is always great to comprehend a new language.}
\item {\em It was a good course that has allowed me to learn a type of programing.}
\end{itemize}

They also found the ASP activities easy and interesting to work on, even if there was a small learning curve: 
\begin{itemize}
\item {\em Once we got into actually working with the program, it was very interesting and I enjoyed playing around with it and figuring out just how simple it really was.}
\item {\em The course structure and the SPARC program make it relatively easy for someone who has not spent much time doing computer science to understand the program and how to program...Following the "methodology" also makes it very easy to learn the program.}
\item {\em It was a bit easier to understand and even enjoy a little after a while, for the TAs went around and took the time to make sure it made sense.}
\item {\em It was a little difficult at the beginning  because I had no prior knowledge of programing. But it was a very enjoyable course and i would take it again.}
\end{itemize}

The main critiques of the course included the need for more specific instructions and expectations for the assignments, better pacing as some participants remarked that the teaching was occasionally too slow or too fast, more time for assignments. Most of these critiques had to do with the delivery of the course. Aside from one comment on the 
ASPIDE programming environment for SPARC, 
%actual workflow of within the SPARC environment, 
there were no explicit critiques of the logic programming paradigm.

\section{Interview Results}
A grounded theory methodology was adopted to generate a theory that explained how students understood and approached computing through ASP activities based on the interview data. Details of the methodology can be found in \cite{corbin2008basics}. This qualitative approach generates a theory or explanation through systematic analysis of data. Grounded theory goes beyond describing and categorizing by finding the processes and relationships in the phenomenon being studied. This approach was conducted on the clinical interview data and served as the main analysis for this study. The unit of analysis was an utterance, which consisted of a complete thought or phrases spoken by the participant. In cases where utterances were rich with many potential units, they were broken down into smaller utterances. Utterances made by the researcher provided context to those made by the participant, but were not analyzed. There were 1,452 utterances used for analysis. 

The first author analyzed the data through an open-coding process coding each utterance with a descriptive label. This process required several iterations to ensure that coding was consistent across the entire dataset. Then, the third author coded the data using the set of labels created by the first author. Through that process, the researchers debated the labels and coding of the utterances, which led to several iterations of re-coding the data and revision of the initial labels. Memos were kept during this analysis process regarding the rationale behind labels and the formation of coding. Since the third author was the expert in the field of ASP and DP and taught the course, his coding was used for the remaining analysis. The resultant set of labels found in open coding are presented in Table~\ref{codes} with their frequency.

\begin{table}
  \caption{Resultant labels from open coding}
  \label{codes}
    \begin{tabular}{l r|l r}
      \hline\hline
Knowledge representation & 137 & Sorts & 28\\
English definition & 131 & Predicate declaration & 26\\
Reasoning & 109 & Commenting & 22\\
Debugging & 86 & Relationship & 19\\
Testing & 71 & Query & 18\\
Rule & 66 & Thinking & 17\\
Syntax & 63 & Translation & 12\\
Strategy & 60 & Reusing & 10\\
Refining & 59 & Prior knowledge & 6\\
Problem understanding & 44 & Reflection & 5\\
Problem solving & 35 & Variable & 4\\
      \hline\hline
    \end{tabular}
    \vspace{-2\baselineskip}
\end{table}

The next step was creating categories that represented larger constructs that were happening in the data. Many of the merged labels were found to relate to each other as well with infrequent labels, which helped form these categories. The five major categories were: Abstraction, Representation, Reasoning, Revision, and Procedures. 
Axial coding analyzed the way these categories were related based on their properties; 
% \cite{corbin2008basics}; 
that is, this part of the analysis found the relationships between the major categories. The first author led the first stage of axial coding which led to these five major categories. Then, these were also debated and discussed by the researchers. The surveys, particularly the post-surveys, were used to guide the axial coding and the creation of the categories. The definitions of the categories also went through several revisions. 

The next sections define each category and give example quotes.

\subsection{Abstraction}
This category describes instances where the participants understand the problem space in more general or abstract terms and are thinking on a higher level within the context of the code they are writing. This type of thought is in contrast to only discussing idea in concrete terms or with specific examples. In this generalization, participants are generally describing relationships with respect to these real-world terms and/or integrating their prior knowledge of the problem space. 

Example quotes included: \textit{The descendant part because the descendants can be anything that's…that was before that person, well before X in this case} (P8);  \textit{So, father of person and person, ok, parent, sister, same parent, gender} (P10);  \textit{I am thinking of like a family tree I guess sort of and so I am thinking of X being or I am thinking of Y being so I am thinking of Y being a person, whose descendants we are looking at…} (P12).  {\em P03: Yeah um I am just trying to think of like a more general way of putting it instead of just listing it all down. (pop up appears and then is closed out) Parent XY and uh… } (P3).

Thus, Abstraction occurs when participants are able to explain and apply the LP concepts within the ASP environment as well as outside it using real-world examples. That is, students' understanding of the LP concepts is not limited to only working within the ASP environment.
\subsection{Representation}
This category refers to the process in which the participants are trying to understand the problem space in coding terms; that is, how their abstraction or understanding of the problem translates into the code. This category includes references to both the conceptual understanding and coding as well as the transition process between the two. 

Example quotes included:  \textit{You can't say, X is the parent of Y if X, if Y is the son, because it's flipped. I hope I am explaining it right. So then I'm saying X is the son of Y if Y is the parent of X, and I'm putting gender because if you are a son you have to be male. And if X is a male. All right I am going to move on to aunt} (P1); \textit{A descendant is like it's the same as a child it's umm like the child of two parents. That is their descendant that's their heir so that is what I am trying to say but I don't know how to say it} (P14); \textit{I put X is the son of Y. I put that in the predicates. This is it. The son is a person. And I know that we have the example here. I can put Matthew is the son of Jon…} (P15) .

Representation is when participants map their understanding of the LP concepts into code; it is a translation between their conceptual understanding and coding. Rather than simplifying this process just as coding, Representation refers to the participants' process in creating a code-based representation of their understanding of the ideas. 
\subsection{Reasoning}

This category is the problem-solving process. It describes instances in which participants were actively thinking about the problem, apply problem-solving strategies associated with ASP, and adapting existing code to solve the current problem. Reasoning was selected as the category name as that was the most frequently used label related to the problem-solving process. 
%Reasoning was selected for the category name as that the most frequent related label. 
%(??sth is missed in the previous sentence??) 
Example quotes included: \textit{So grandchild Isaac of George, grandchild Joseph, George, grandchild Susie, George. Grandchild, those are the children, so, grandchild, grandchild, X is a grandchild of Y, if Y is grandfather, if Y is the grandfather of X and X is the, do we define gender? Yes, gender of X is male, because grandfather has to be a male} (P2); 
% \textit{I have like separate ones like I have sister and I have like aunt and I have parent} (P5); 
 {\em I think I’m doing aunt. (Begins typing comment for aunt.) ...... For any person X and person Y, X is the parent of Y. There is a person Z, Z is the sister of X such that X is the sister of Z… I mean Z is the sister of X and the aunt of Y.} (P5).
 \textit{Nephew, so then you have Isaac, and actually you can create a rule} (P11).

Thus, Reasoning describes the entire problem-solving process as participants attempt to negotiate their understanding of the problem and the solution. It was this reasoning process -- the internal discussions within themselves and the reflection -- that was the most observed activity in the problem solving process, which is why this category was named Reasoning.  

\subsection{Revision}
Revision encapsulates a lot of the overall debugging process, but emphasizes the importance of participants asking questions in SPARC to see if their code was correct. This category includes the participants' process of questioning to see if they completed the tasks correctly, which was done through querying to see if their solutions were right.
Example quotes included: \textit{So I'll write aunt and then I'll write Rose comma  Susie, % \todo{check Rose comma Susie?}  
and then I'll have to add a question mark, and then have to press execute} (P1); \textit{Ok and now I need to write queries to see if my program works} (P14); \textit{I am going to ask who are descendants of George because George is like the patriarch of the family} (P12). Simply put, Revision is the debugging process after the participants developed their initial solution.

\subsection{Procedures}
This category represents a large part of the data in which participants were either restating or interpreting facts or instructions. In some cases, participants were asking if they had the correct interpretation of the assignment instructions. In this category, participants frequently mentioned specific rules and strategies associated with declarative programing. Often, these utterances were co-labeled with other more descriptive categories such as Representation and Reasoning. 
Example quotes included: 
% \textit{Whenever I am talking about the son I forgot to include that it has- and he has to be male, and that X has to be male} (P1); 
{\em So I write. I have to comment first do that the people, errr, so that anyone else that is reading it knows what I am trying to say. 
...... So I said that son X and Y denotes. Hmmm! It’s all caps (small laugh). Denotes that X is the son of Y. And then I write the actual code and just put son,  I put in parenthesis, uhh, ......} (P1); 
{\em OK so X is the son of Y if Y is the parent of X and X is male. So… Gender um, Oh I was reading the English,  ......} (P3);
\textit{So now I am at son. Oh forgot to add the periods. Ok. Got it, got it, and got it. Now I need it aunt} (P8); \textit{Jon is the father of Matthew, so that rule that I wrote works} (P16).
So, in the case of ASP, the procedures category describes the participants' actions and ability in identifying the instructions and facts involved with the problem, which eventually helped them to develop a solution for their problem sets.

\subsection{Order of Category Occurrences}
\begin{figure}
\includegraphics[scale=.6]{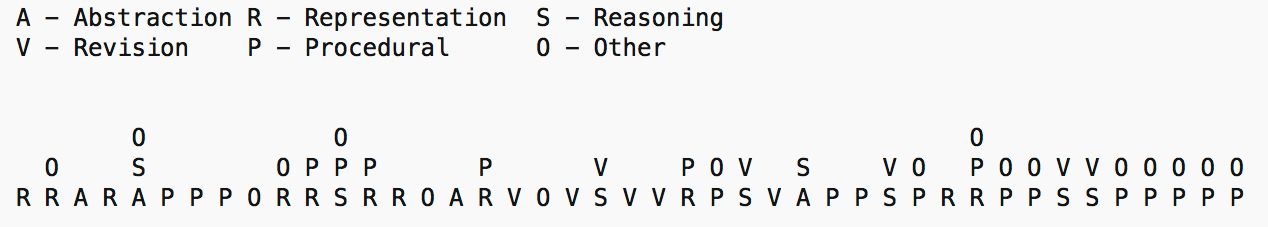}
\caption{Sample (Participant 12) of order of category occurrences during clinical interview}\label{fig:graph}
\end{figure}

As part of axial coding, there was an analysis of when these categories occurred during the clinical interview to learn more about the categories themselves and how they related to the other categories. A letter-coded graph plotted out the occurrence of each category in the order they appear with each participant.  Figure~\ref{fig:graph} shows an example of this graph. When an utterance had more than one category associated with it, plots are stacked. 

%% (??can the current graph show length--not on the current graph, but I'll remove the line Plots varied in length due to the variance in number of coded utterances.??)

A visual analysis of all participants' graphs was conducted to describe when each category occurred and the relationships between those occurrences. Across all participants, interviewers had them all read the instructions once at the beginning, but these utterances were not coded. Since participants were asked to work on their assignments, some participants had already started on them prior to the interview. Thus, those graphs immediately started with non-procedural labels. Findings from a visual analysis of all participants graphs are presented in the following subsections.
%(?? full stop or the findings from the visual analysis are presented in the following subsections??)

\subsubsection{Abstraction} 
Evidence of Abstraction mostly came from the beginning to the middle for most participants. For several participants, it occurred alongside Procedures (P3, P7, P8, P10, P11, P12, P13) as well as with Representation (P4, P9, P11, P12, P14, P16). Three participants had Abstraction after Revision, which may suggest that the questioning process may support part of abstraction abilities. Another three participants (P2, P3, P13) showed no evidence of Abstraction during the interview. However, the interviewer reported that P2 did not get a chance to test their program and P3 had technical issues with the computer. P13 was reported to feel confident about what they were doing, and started Revision early.
%(??their ok here, yes??)
% (??computer - computer is fine??)(.

\subsubsection{Representation}

Similar to Reasoning and Procedures, Representation also happened throughout the interview. For the most part, Representation started before clusters of Revision for 11 participants (P1, P3, P4, P5, P7, P8, P9, P11, P12, P15, P16). However, for three of these participants (P2, P11, P16), there were more revision afterward. For P13, Representation was intertwined into Revision throughout the interview. Most of P14's interview consisted of utterances labeled as Representation as he was mostly listing out what had to be represented in the knowledge base. 

\subsubsection{Reasoning} 
Reasoning was noted throughout the interviews for most participants, similar to the procedural utterances, which may suggest that it is an important recurring process. Reasoning occurred with and between instances of Revision (P1, P16) and Representation (P2, P13), Abstraction (P14, P15). Thus, Reasoning is also intertwined with other parts of the ASP process. There was no evidence of Reasoning from P6; however, there were not many coded utterances from that participant in general.

\subsubsection{Revision}

Most participants (P1, P2, P3, P4, P6, P7, P8, P9, P11, P15) were engaged in revision during the last half the interview, which was expected as they spent the first half identifying the objects, relationships, knowledge, and coding before testing. P10 did not get as far during the interview to be engaged in revision. Other participants had already started their assignments before the interview, so they began to revise their solutions earlier than others. 

\subsubsection{Procedures}

All participants had instances of utterances of procedure, which were distributed throughout the interview as well as simultaneously appearing with other categories. An interesting observation was that there were procedural moments throughout the interview, but not just at the beginning. That is, participants kept going back to the instructions and/or the facts presented in the assignment. 

At first, the Procedures category did not seem to provide much insight into how the participants approached computing and computer science. After all, these utterances were of participants reading and re-reading the rules and asking if they understood the instructions and information correctly. Procedural utterances were made throughout the entire computing process. In selective coding, analysis examined how one category could be connected to other categories \cite{corbin2008basics}: the Procedures category was found as that central theme. Identifying objects and relations in the problem, through English definitions, by carefully reading the problem description and integrating one's common sense knowledge was a major component of the explicit methodology for solving LP problems. Thus, there was deep interaction between the Procedures and other processes. Students had also mentioned in their post surveys that they had learned this methodology to solve the problems. Although that may not be surprising, it showed that they were able to adopt those strategies, which in turn, led to more reasoning and representation processes. 

\section{Achievement Results}

%(??does subsections 5.1 - 5.3 fit under this section title--you are right, I will fix and renumber??)

%(??checking 2.3.3 to make sure discussion here is consistent with that there??) (??move this "subsection" on lab grades and explanation after the section 5.1 on model??) 

Each lab was scored using a rubric, which had five parts: completion of requirements, abstraction, correctness, coding style/readability, and documentation (commenting). Each component was scored on a scale from 0 (poor) to 4 (excellent) by the grader. Table~\ref{scores} shows the scores and completion rate for Labs 2 through 4. 
Due to the logistics issues, we were not able to collect the submissions of Lab 1.

\begin{table}
  \caption{Average lab scores rounded to the nearest whole number: COMPLETE = completion, ABS = abstraction, CORRECT = correctness, STYLE = coding style, DOC = documentation, TOT = total}
  \label{scores}
    \begin{tabular}{lcccccc}
      \hline\hline
     \textbf{Lab \#} & \textbf{COMPLETE} & \textbf{ABS} & \textbf{CORRECT} & \textbf{STYLE} & \textbf{DOC} & \textbf{TOT}\\
      \hline
	Lab 2 (N=16) & 95\% & 88\% & 82\% & 100\% & 84\% & 90\%\\
	Lab 3 (N=14) &100\% & 92\% & 92\% & 95\% & 70\% & 90\%\\
	Lab 4 (N=11) & 89\% & 98\% & 73\% & 100\% & 96\% & 91\%\\
	All labs & 95\% & 92\%  & 83\% & 98\% & 82\% & 90\%\\
      \hline\hline
    \end{tabular}
\end{table}

Overall, participants did well on all three labs across each component. The average score of completion was 95\% (3.8 out of 4 points). This high average shows they were able to address the project requirements as well as follow the two-step methodology.  The average score for abstraction was 91.9\% (3.68 out of 4 points).
 The average score for correctness (e.g., getting the correct output) was lower than completion and abstraction: 83.1\% (3.32 out of 4 points).  For 
%correctness and 
completion, the grader looked for attempts made by students to address each requirement. Thus, a participant could have made a reasonable attempt at all the requirements, but implementation did not fully match the specifications or yield correct results for all test cases.  Similarly, abstraction scores would also be lower in those events. With respect to the model %(??our model for students to approach LP--it's implied??) 
 in Figure~\ref{fig:model}, this quantitative data supports the notion that students are able to engage in Representation (coding), Abstraction, and Reasoning (problem solving) within the LP paradigm. The lower correctness scores may demonstrate that the participants may need to improve in the area of coding. 
 % \todo{what do you think?} 
The ordering of the categories showed that revision did not happen until the second half of the interview, if at all, which places some limitations on the revision (debugging) process. The high completeness score is also in line with the dominance of the Procedures category during the task, corresponding to the predictions of our model, since participants mostly attempted to address all the requirements. Coding style was exceptionally good across all the labs at 98.1\% (3.92 out of 4 points). The lowest component was documenting code, which was 81.9\% across the labs (3.28 out of 4 points). Generally, most students that submitted the assignment received high scores: the average lab score was 90\% (18 out of 20 total points). 
The lower submitted rates in the later labs was attributed to student absences mainly resulted from family vacations (which are not rare in 
the summer in the United States).
%The lower submitted rates in the later labs was attributed to student attrition: students dropped out (i.e, some students left for family vacations) throughout the course. 

\section{Model of How Participants Approach Computing through LP}

\begin{figure}
\includegraphics[scale=.5]{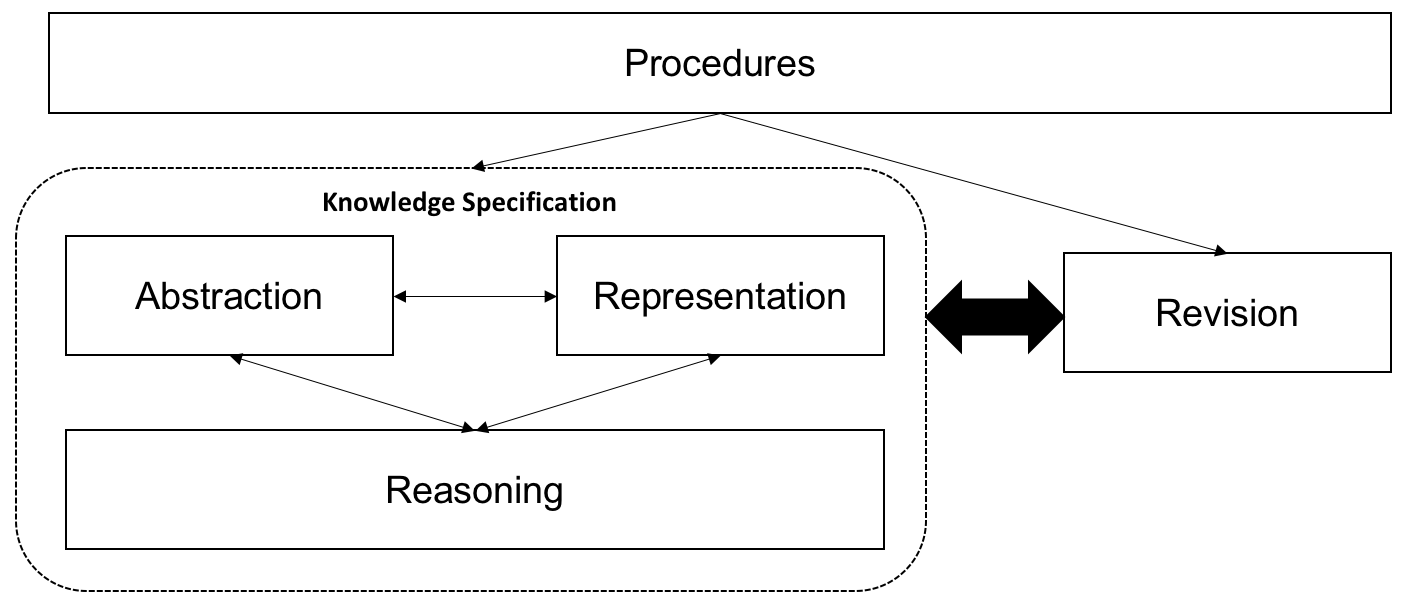}
\caption{Model of how participants approach Logic Programming through Answer Set Programming}\label{fig:model}
\end{figure}

Based on the grounded theory analysis, supported by the visual analysis of ordering of categories, a model was constructed that may explain how participants approach computing through LP as based on the data. 
It is shown in Figure~\ref{fig:model}.   
This figure highlights the foundational importance of following the strategies (i.e., explicit modeling methodology) of declarative programming. 
In the case of Answer Set Programming, it allows for an explicit methodology, which guides students towards abstracting the concepts, representing it in code, and reasoning in their problem solving.  Abstraction, Representation, and Reasoning 
%(??capital letters are intentional--yes, the reviewer suggested it??) (??it seems to be natural to put this model section after section 4.6.--we need some description here. ??) 
often occurred together and are grouped separately from Revision, %(??Revision if your decide all category name should be capitalized. Once decision is made, make sure to check all names. I didn't exhaust comments on this issue in the rest of the paper??)
which mostly came toward the end of the task. Similarly, Abstraction  and Representation happened more sequentially with Abstraction coming before Representation. Reasoning was observed throughout the interview, which is indicated by a larger rectangle in the figure.

The model shows  the importance of Procedures in logic programming; in this case, students were able to follow the explicit methodology in LP. The students also demonstrated the general problem solving skill of iterative refinement. 
There are two distinguishable big loop components: Knowledge Specification and Revision. The knowledge Specification component is again an iterative refinement consisting of interleaving steps: Abstraction, Representation and Reasoning. As an example, when a student tries to specify their knowledge about aunt, they try to figure out the objects (individual persons) and relations (e.g., who is whose sister, or who is whose parent) and then tried to define their knowledge about aunt using these objects and relations. Participants tried to make sure the definition captures their intended meaning, which requires them to reason with and understand the relations better. In fact, this model reflects the intended skills most researchers would like the students to obtain in an introductory course: the skills should be general but important in computing (and beyond). The skills include: the understanding and the application of the LP procedures (a general methodology for problem solving), iterative refinement in problem solving, programming, abstraction, rigorous high level (logic) representation of knowledge and (logic-based) reasoning with the knowledge. 

Logic Programming can be thought of as allowing for an explicit problem solving methodology: problem understanding, precise representation, and reasoning with knowledge. This methodology is almost universal in problem solving. It will particularly be useful as LP lays the foundation for problem understanding and knowledge representation. This foundation is expected to be helpful when students study algorithms in future.
% and thus, focus on algorithmic elements. 
%The gap between problem specification and logic programming constructs (syntax/semantics) is smaller. (?? remove the left sentence??) 
% In Logic Programming, problem specification coding, in imperative: problem specification – figure out how – coding. \todo{This last sentence is incomplete}

\section{Discussion}

The research question asked how using LP, specifically through ASP, to teach an introductory computer science course for high school students impacts their understanding of computer science and computing. 

Prior to this course, most participants had limited experiences with CS, and certainly even more limited experience with declarative programming. Their perception of CS was mainly connected to programming and creating technologies for the world. Given that participants were unfamiliar with declarative programming, they attempted to explain within the context of a paradigm with which they were familiar: imperative and, to an extent, object-oriented programming, because of their previous courses in Alice and Scratch. Interestingly, after this course, they were able to explain the tasks and procedures, 
%they did 
even though they were unable to articulate what declarative programming was. The data suggests that students may not even understand what programming paradigms are. 

The students had done very well in their lab assignments with an average score of 90\% overall on all labs. Students also reported on post-surveys that they were able to learn how to program with SPARC and follow the modeling methodology in their computing activities. Such results are in line with other research that shows that LP can be easy to learn \cite{mendelsohn1990programming}. The main reason is the simple syntax and intuitive/natural semantics of ASP which allow the students to focus less on language specific feature and more on the problem solving skills. The clinical interview data shows further that the students were able to learn and apply the important concepts underlying computing. The model (Figure~\ref{fig:model}) shows that they were able to apply explicit LP methodologies in problem solving and iterative refinement in both big steps (Knowledge Specification and Revision) and small steps (Abstraction, Representation and Reasoning), when solving the problems. Participants applied their knowledge and skills in Abstraction, Representation and Reasoning heavily, in an interwoven manner, in understanding, extracting and defining the knowledge needed in solving a given problem. Once they identified/defined the knowledge needed, they are able to carry out the standard programming tasks from coding to debugging.

The overall scores for completion (95\%) showed that students were able to make reasonable attempts at meeting lab requirements. Though students were able to follow the two-step methodology, the lower correctness does bring attention to a need to help students address all specifications and provide more support during the revision process. The lower documentation score (81\%) also suggests that students need to be encouraged to document their code. However, the instructional staff did refer to time being a limitation, which may explain higher completion, abstraction, and correctness (slightly) scores over documentation: that is, students may have focused on getting the project done and making it work first. This study shows that working within the LP paradigm yields similar results as within a traditional imperative or object-oriented paradigm. The model generated shows that students can engage in computing skills, such as abstraction, problem solving (Representation), and debugging (Revision). These important LP skills are aligned with four of the six core computational thinking practices found in the AP CS Principles Framework \cite{apcs}: connecting computing, creating computational artifacts, abstracting, analyzing problems, and communicating, and collaborating. %, (??communicating, and collaborating??). 
%(??in our new introduction, we mentioned four components shared by standards: abstracting, algorithm, programming and communicating. In the sentences above, should we simply connect back to those four aspects, instead of the AP course only--I think we should just simply connect it back as the connections are implied. We were explicit in the intro. ??)

Similarly, the study shows that participants were always engaged in cyclical process of Abstraction, Reasoning, and Representation. In fact, LP requires programmers to have a solid grasp of the problem space before starting on the solution. The data also shows that participants were engaged in abstraction -- they were able to think about the major ideas in a more generalized context.   In this course, they were also expected to explain and connect basic computing concepts. Indeed, the participants were able to explain computing concepts within the context of this course and tasks. Students were able to operate within the LP paradigm driven by the explicit methodology for LP; however, they may not immediately see how it is applicable outside the context of the assignments. But, they understood the LP concepts and processes and were successful in completing their assignments. More work needs to be done to make connections to how they see CS in the real world through LP: students understood the problem-solving nature of CS and believed LP was one way of solving problems. 

\subsection{Limitations}
\label{sec:limitations}
Participants were above average students and were pre-screened for this program, and had some experience with Scratch and Alice. However, participants stated no prior experience with LP or Declarative Programming. 
%Though every student initially volunteered for this study, 
% there were still some participants who missed some stopped attending or did not complete an assignment. For example, some participants left the course for vacations. 
A few students missed some assignments mainly due to family vacations. 
Lastly, this course met for four weeks, which provided  limited exposure to LP as compared to regular secondary school or university courses. Future studies on LP in introductory CS courses could be conducted on full semester courses rather than short courses. 

%(??Replace the left sentence by: during the class, some students missed some sessions mainly due to summer family vacations.??) 

Also, it is worth of noting that {\em algorithms} are also taken as an important component of Computer Science in K-12
education.  As a pure declarative programming 
paradigm, ASP is hardly the best option for teaching 
algorithms. In our perspective, one reasonable sequence of computer science courses would be teaching the foundations (except algorithms) of CS using ASP, and introducing algorithms in a subsequent course.\footnote{If one really wants to introduce algorithmic ideas in the foundation class, one can employ the unplugged approach \cite{feaster2011teaching} which is
independent of a programming language.} 
The advantage of this approach is to shorten the distance from problem understanding and representation to programming by skipping the algorithm component which is certainly non-trivial and significantly prolongs the learning process before students can solve interesting problems using computers. This advantage becomes more prominent given that the computer science education in K-12 is expected to serve all students. 

We do note that ASP can cover some  important aspect(s) of algorithms. One example is recursion in ASP. It is also argued that a precise problem specification is desirable before the study of algorithms solving them  \cite{kleinberg2006algorithm}. As shown in our data, 
LP based modeling helps to develop students' capacity on abstraction and representation (in a rigorous manner) which are key skills in precisely specifying problems. 

\section{Conclusions}

This paper asserts that CS educators should take a closer look at using LP during the introductory courses. As suggested in \cite{ball2015teach}, future computer scientists should be equipped with foundational programming language principles involving logic and formal specification to design and implement complex software systems needed by the society. Our results show that the students were able to focus on the key concepts of computing including abstraction, representation and reasoning when solving problems. The results show evidence that it is viable to teach an introductory computing course using ASP. % (or Logic Programming). 
It may also bring attention to how early we can teach students about different paradigms and if that would expand the students' conceptions of computing in order to engage in abstraction and problem solving within the declarative programming paradigm. This effort would later require educators to build the connections between these different paradigms and the rest of CS.

%although they may not understand the bigger picture of how it fit into CS. (?? the left sentence is pretty complex. are you able to break it??) 

\section{Acknowledgments}
The authors thank Cynthia Perez, Rocky Upchurch and Edward Wertz for their contributions to this project, and thank Michael Gelfond for valuable discussions and sharing his teaching materials. This work is partially supported by NSF grant CNS-1359359. We thank the 
anonymous reviewers whose feedback helps to improve the quality of this paper.

%\bibliographystyle{acmtrans}
%\bibliography{new_tlp2egui}

\label{lastpage}
\end{document}